\documentclass[aps,prb,reprint,twocolumn,unsortedaddress,showpacs,amssymb]{revtex4}
\usepackage[english]{babel}
\usepackage{graphics}
\usepackage{graphicx}
\usepackage{epsfig}
\usepackage{amssymb}
\usepackage{dcolumn}
\usepackage{bm}
\usepackage{color}
\usepackage{natbib}
\usepackage{lineno}
\usepackage{sidecap}
\usepackage{verbatim}  
\usepackage{vector}  
\usepackage{wrapfig}
\usepackage{enumerate}
\usepackage{sidecap}
\usepackage{amsmath}
\usepackage{hyperref}

\begin{document}

\title{Electron correlation effects and scattering rates in Fe$_{1+y}$Te$_{1-x}$Se$_x$ superconductors}

\author{S.\ Thirupathaiah}
\email{t.setti@sscu.iisc.ernet.in}
\affiliation{Solid State and Structural Chemistry Unit, Indian Institute of Science, Bangalore, Karnataka, 560012, India.}
\author{J.\ Fink}
\affiliation{Leibniz-Institute for Solid State and Materials Research Dresden, P.O.Box 270116, D-01171 Dresden, Germany.}
\author{P. K.\ Maheswari, V. P. S. Awana}
\affiliation{CSIR-National Physical Laboratory, New Delhi 110012, India.}
\author{E.\ Slooten, Y.\ Huang, M.S.\ Golden}
\affiliation{Van der Waals-Zeeman Institute, IoP, University of Amsterdam, NL-1098 XH, Amsterdam, The Netherlands.}
\author{F.\ Lochner}
\affiliation{Institute f$\ddot{u}$r Theoretische Physik III, Ruhr-Universit$\ddot{a}$t Bochum, D-44801 Bochum, Germany.}
\author{R. Ovsyannikov}
\affiliation{Helmholtz-Zentrum Berlin, Albert-Eistein-Str. 15,  D-12489 Berlin, Germany.}
\author{H. D\"urr}
\affiliation{Stanford Institute for Materials and Energy Sciences, SLAC National Accelerator Laboratory, 2575 Sand Hill Road, Menlo Park, CA94025, USA} 
\affiliation{Van der Waals-Zeeman Institute, IoP, University of Amsterdam, NL-1098 XH, Amsterdam, The Netherlands} 
\author{I. Eremin}
\affiliation{Institute f$\ddot{u}$r Theoretische Physik III, Ruhr-Universit$\ddot{a}$t Bochum, D-44801 Bochum, Germany.}
\affiliation{Institute of Physics, Kazan (Volga Region) Federal University, 420008 Kazan, Russian Federation.}

\date{\today}

\begin{abstract}
Using angle-resolved photoemission spectroscopy we have studied the low-energy electronic structure and the Fermi surface topology of Fe$_{1+y}$Te$_{1-x}$Se$_x$ superconductors. Similar to the known iron pnictides we observe hole pockets at the center and electron pockets at the corner of the Brillouin zone (BZ). However, on a finer level, the electronic structure around the $\Gamma$- and $Z$-points in $k$-space is substantially different from other iron pnictides, in that we observe two hole pockets at the $\Gamma$-point, and more interestingly only one hole pocket is seen at the $Z$-point, whereas in $1111$-, $111$-, and $122$-type compounds, three hole pockets could be readily found at the zone center. Another major difference noted in the Fe$_{1+y}$Te$_{1-x}$Se$_x$ superconductors is that the top of innermost hole-like band moves away from the Fermi level to higher binding energy on going from $\Gamma$ to $Z$, quite opposite to the iron pnictides. The polarization dependence of the observed features was used to aid the attribution of the orbital character of the observed bands. Photon energy dependent measurements suggest a weak $k_z$ dispersion for the outer hole pocket and a moderate $k_z$ dispersion for the inner hole pocket.  By evaluating the momentum and energy dependent spectral widths,  the single-particle self-energy was extracted and interestingly this shows a pronounced non-Fermi liquid behaviour for these compounds. The experimental observations are discussed in context of electronic band structure calculations and models for the self-energy such as the spin-fermion model and the marginal-Fermi-liquid.
\end{abstract}

\maketitle 

\section{Introduction}
The present consensus for the normal state of the high-$T_c$ iron-based superconductors is that they show strange metallic character~\cite{Analytis2014, Meingast2012} near a quantum critical point (QCP)~\cite{Dai2009a}, that is reached by either charge carrier doping, chemical pressure or by applying mechanical pressure to the parent compound.~\cite{Rotter2008,Sefat2008,Jeevan2011} This strange metallic character is attributed to strong antiferromagnetic spin fluctuations, originating from interband scattering between the hole and electron pockets located in the center and corner of the Brillouin zone, respectively.~\cite{Mazin2008a} Comparing the iron pnictides and the iron chalcogenide systems, FeTe and FeSe, the latter have been suggested to possess stronger many-body correlation effects near the Fermi level from density functional theory (DFT) plus dynamic mean-field theory (DMFT) calculations.~\cite{Kotliar2006} This conclusion is supported further by transport measurements~\cite{Noji2012} and photoemission~\cite{Tamai2010,Yamasaki2010,Maletz2014,Lubashevsky2012} experiments. More recent theory work has also argued that for the iron chalcogenide systems, electron correlations lead to bad-metal behavior, despite the intermediate values of the Hubbard repulsion U and Hund's rule coupling J.~\cite{Lanata2013}

Strong interest in the iron chalcogenides has been rekindled recently due to the spectroscopic observation of superconducting energy gaps at and above the boiling point of liquid nitrogen for single unit-cell thin films of FeSe on SrTiO$_3$ substrates.~\cite{Qing-Yan2012} These systems are now the record-holders for highest $T_c$ in the Fe-based superconductors. Recently, ARPES data have been modelled to extract theoretical parameters suggesting that coupling of a SrTiO$_3$ phonon can significantly enhance the magnetism-driven pairing energy for the electrons in the single unit cell thick film of FeSe.~\cite{Lee2014}    

Also of interest have been the recent and ongoing discussions as to whether a Fermi liquid ground state is the appropriate description for optimally n-type (electron) doped BaFe$_2$As$_2$. DMFT calculations argue for canonical Fermi liquid character when the Ba122 compound is optimally doped with electrons, while optimal hole doping leads to strong band renormalization near the Fermi level and thus to non-Fermi liquid character.~\cite{Werner2012} Recent optical experiments on n-doped BaFe$_2$As$_2$ would seem to offer partial support for this,~\cite{Tytarenko2015} but other experimental data from transport,~\cite{Kasahara2010} thermal properties,~\cite{Meingast2012} NMR,~\cite{Ning2010a} quantum oscillations~\cite{Shishido2010} and photoemission measurements~\cite{Fink2015} suggest non-Fermi liquid character near the quantum critical point in the BaFe$_2$As$_2$ system when doped with charge carriers of either sign or upon applying chemical pressure. 

Thus, given the backdrop of new data and insights into novel, high temperature pairing phenomena in the iron chalcogenides, and the ongoing, lively discussions as to the Fermi liquid (or not) behavior in the iron pnictides, it is of great interest to examine the iron chalcogenides from the point of view of Fermi liquid theory and how strong electron correlations make themselves felt. Indeed, one theory report suggests non-Fermi liquid behaviour also for the chalcogenides.~\cite{Aichhorn2010}

There are various angle-resolved photoelectron spectroscopy (ARPES) studies~\cite{Tamai2010, Xia2009a, Zhang2010f, Lubashevsky2012, Maletz2014, He2013,Starowicz2013, Liu2012, Tan2013, Xia2009a} which indicate that the Fe chalcogenides show strong electron correlation effects. It should be noted that all these studies of correlation effects have been carried out in the $\Gamma-M-X$ plane in 3D $k$-space. As yet, no report has been made of if and how the picture changes upon variation of the $k_z$ value in these compounds, and most experimental studies have inferred the impact of electronic correlation from the renormalisation of the band structure (band velocity).    

In this paper we present electronic structure studies of Fe$_{1+y}$Te$_{1-x}$Se$_x$ superconductors using and combination angle-resolved photoelectron spectroscopy (ARPES) and DFT calculations.
We compare the experimental results with our DFT calculations, as well as with other existing experimental and theoretical reports on these systems.~\cite{Singh2010a,Tamai2010,Chen2010d,Subedi2008c, Chen2010a,Nakayama2010,Mizuguchi2009,Margadonna2009,Medvedev2009,Aichhorn2010}
Our ARPES data enable attribution of the orbital character of the bands involved (by exploiting photon polarization) and we explicitly examine the role of $k_z$ (by variation of the photon energy) for the hole pocket states along the $\Gamma-Z$ ($k$) direction. The data suggest weak $k_z$ dispersion for one hole pocket, while a moderate $k_z$ dispersion is observed for the other hole pocket at the Brillouin zone center, a result which is consistent with the DFT calculations.
In agreement with previous reports,~\cite{Xia2009a,Liu2015} the hole pockets display a mass renormalization ($m^*/m_b$) of 2-4 at higher binding energies. The experimental data are also analyzed with respect to a possible $k_z$ dependence of the mass renormalization and the Fermi velocity ($v_F$). 
The ARPES data have also been fitted so as to enable estimation of the imaginary part of the self-energy ($\Im\Sigma$). Our results suggest a departure from canonical Fermi-liquid behaviour for the quasi-particles near the zone center.
In particular, the imaginary part of the self-energy is linear in energy for the inner hole pocket, whose band top generates a van Hove singularity (vHs) near the Fermi level. 
This linear-in-energy self-energy is shown to be well described using a marginal-Fermi-liquid theory (MFL)~\cite{Varma1989} approach with a coupling constant ($\lambda$) of 1.5.

\section{Experimental details}
ARPES provides information on the energy and momentum dependent spectral function.~\cite{Huefner1994} By detecting the emitted photoelectrons at various angles one can extract the in-plane ($k_x-k_y$ plane) electronic structure, while by changing the photon energy it is possible to derive the $k_z$ dependent electronic structure. Using polarized photons, due to the matrix element effects, it is possible to obtain information on the orbital character of the detected bands.

Single crystals of Fe$_{1.068}$Te$_{1-x}$Se$_x$ ($x$ = 0.36 and 0.46)  were grown in Amsterdam by the Bridgman technique using self-flux. The crystals show superconducting transitions at $T_c$ $\approx$ 11 K and 15 K with $x$=0.36 and 0.46, respectively. Further elemental analysis on these single crystals are reported elsewhere, as are data showing them to possess simple, high quality and non-reconstructed cleavage surfaces.~\cite{Massee2009b}. 
Another set of high quality of Fe$_{1+y}$Te$_{0.5}$Se$_{0.5}$ (y \textless 1$\%$) single crystals were grown in NPL, Delhi using the self-flux growth technique. These crystals showed a $T_c$ of 14 K. The elemental analysis of these crystal is reported elsewhere.~\cite{Maheshwari2015}
  
ARPES measurements were carried out in BESSY II (Helmholtz Zentrum Berlin) synchrotron radiation facility at the UE112-PGM2b beam line using the "1$^3$-ARPES" end station equipped with SCIENTA R4000 analyzer.~\cite{Borisenko2012a,Borisenko2012b} The total energy resolution was set between 5 and 10 meV, depending on the applied photon energy. Samples were cleaved $\textit{in situ}$ at a sample temperature lower than 20 K. All the measurements were carried out at a sample temperature $T\approx$1 K.

\section{Calculations}
To understand the experimental data we have performed a theoretical analysis of the electronic band structure of FeSe, following Ref.~\onlinecite{Eschrig2009}. Using a three-dimensional tight-binding parametrization of the LDA (local density approximation) band structure, we computed the Fermi velocity variation along the $k_z$ direction for the three hole pockets near the center of the Brillouin zone.
The hole pockets which possess mostly ${xz}$ and ${yz}$ character, mixed with ${x^2-y^2}$, demonstrate a weak $k_z$ variation of the Fermi velocity, while the variation is stronger for the hole pocket which has an admixture of $z^2$ orbital character. We expect this variation to be further enhanced by the effects of short-range electronic correlations, which are not included in our LDA-based calculations.

\section{Results}
\subsection{ARPES data: Fermi surfaces and band dispersions}

\begin{figure}
	\centering
		\includegraphics[width=0.48\textwidth]{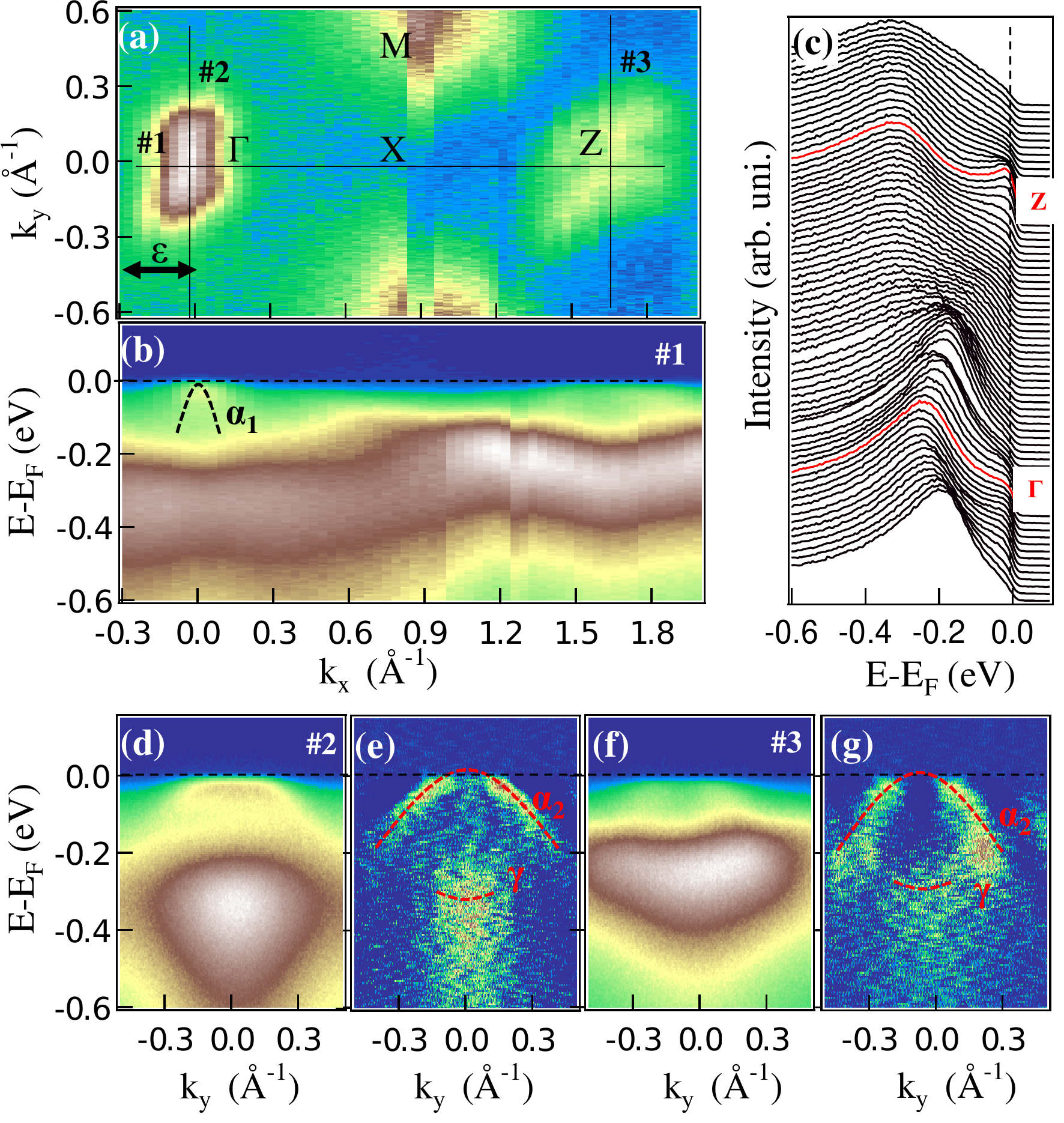}
	\caption{(Color online) ARPES spectra of Fe$_{1.068}$Te$_{0.54}$Se$_{0.46}$ measured with an excitation energy h$\nu$=75 eV using $p$-polarized light. (a) is the Fermi surface map. The light polarization vector ($\vec{\varepsilon}$) is displayed on the figure. Panels (b), (d) and (f) show energy distribution maps (EDMs) taken from cuts \#1, \#2 and \#3, which are overlaid on the Fermi surface map. Panel (c) shows the energy distribution curves (EDCs) from the EDM shown in (b). Panels (e) and (g) contain the second derivatives of the EDMs shown in (d) and (f), respectively. The sample temperature was 1K.}
	\label{1}
\end{figure}

\begin{figure}
	\centering
		\includegraphics[width=0.48\textwidth]{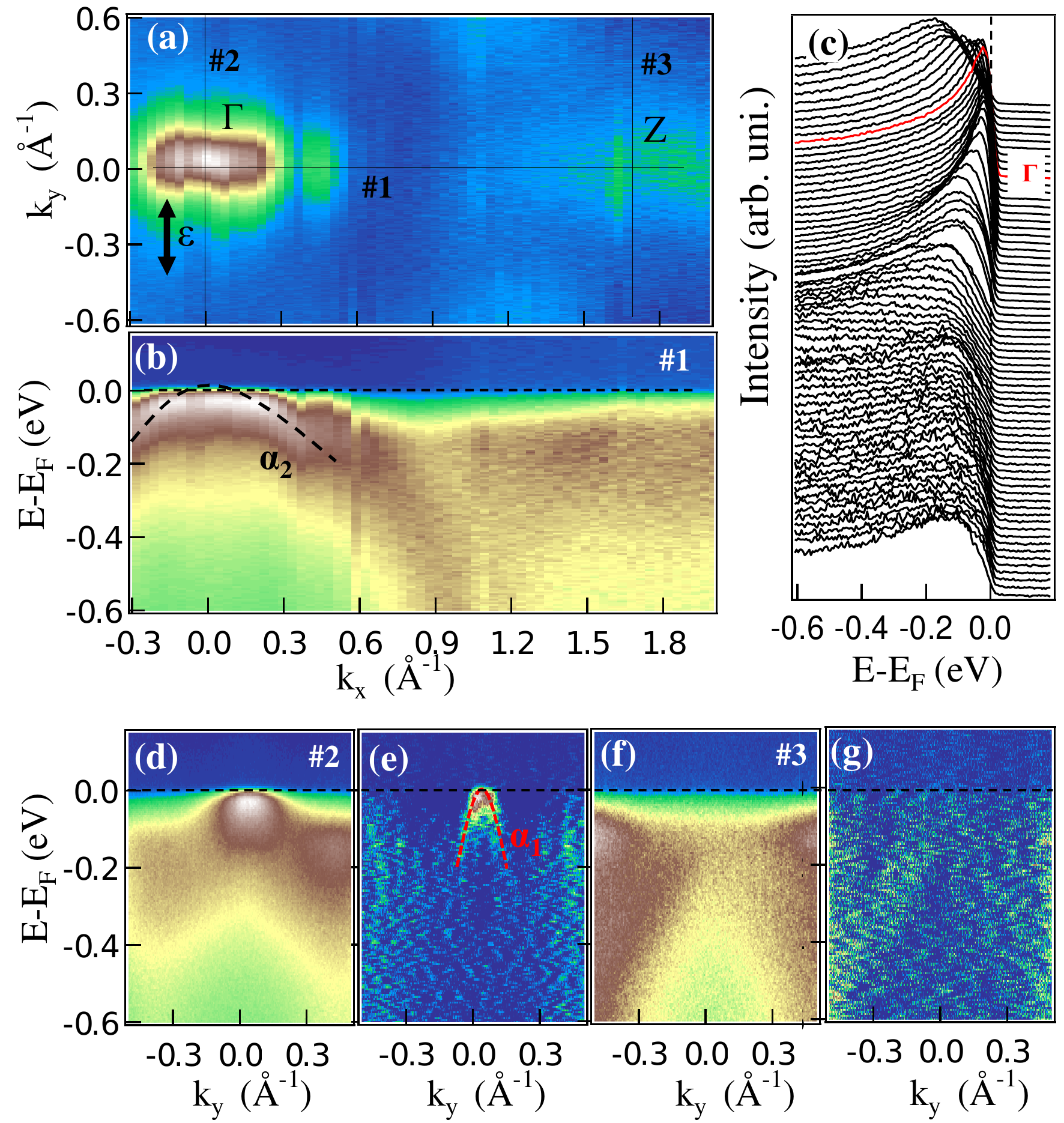}
	\caption{(Color online) ARPES spectra of Fe$_{1.068}$Te$_{0.54}$Se$_{0.46}$ measured with an excitation energy h$\nu$=75 eV using $s$-polarized light. (a) is the Fermi surface map. The light polarization vector ($\vec{\varepsilon}$) is displayed on the figure. Panels (b), (d) and (f) show energy distribution maps (EDMs) taken from the cuts \#1, \#2 and \#3, which are overlaid on the Fermi surface map. Panel (c) shows the energy distribution curves (EDCs) from the EDM shown in (b). Panels (e) and (g) contain the second derivatives of the EDMs shown in (d) and (f), respectively. The sample temperature was 1K.}
	\label{2x}
\end{figure}

Figure~\ref{1} shows the ARPES spectra of the Fe$_{1.068}$Te$_{0.54}$Se$_{0.46}$ superconductor, recorded along the $\Gamma-X$ high symmetry line using $p$-polarized light with an excitation energy h$\nu$=75 eV. The Fermi surface (FS) map shown in Fig.~\ref{1} (a) results from integration over an energy  window of 10 meV centered at the Fermi level ($E_F$). In Fig.~\ref{1} (b), we show an I(k,E) image (EDM or energy distribution map), taken along the cut~\#1 as shown on the FS map in panel (a). Similarly, Figs.~\ref{1} (d) and (f) depict EDMs along cuts~\#2 (through $\Gamma$) and \#3 (through Z), respectively. Fig.~\ref{1} (c) shows energy dispersion curves (EDCs) taken from the EDM shown in Fig.~\ref{1} (b).
The data shown in panels (e) and (g) of Fig.\ref{1} are the second derivative of the EDMs shown in Figs.~\ref{1} (d) and (f), respectively.
The data shown in Figs.~\ref{1} (b-d) clearly show the existence of two hole-like bands, which we label $\alpha_1$ and $\alpha_2$, at the center of the Brillouin zone.
The band $\alpha_1$ disperses strongly towards $E_F$ but does not cross it, forming a van Hove singularity near the Fermi level, consistent with the iron pnictide superconductors.~\cite{Liu2010b, Zabolotnyy2009a, Borisenko2010, Lubashevsky2012, Thirupathaiah2013} The $\alpha_2$ hole-pocket crosses $E_F$ at a Fermi wavevector ($k_F$) of 0.15$\pm$0.02~$\AA^{-1}$.
At $Z$, the high symmetry point is reached at a larger polar angle, and we observe a band having weak spectral weight crossing $E_F$ at a $k_F$=0.16$\pm$0.02~$\AA^{-1}$.  This observation of hole pockets at the zone center is in keeping with previous reports on these compounds.~\cite{Tamai2010,Chen2010d}
Following an analysis of the measurement geometry and polarization dependent selection rules laid out in detail in Ref.~\onlinecite{Fink2009}, it can be concluded that the even parity ${xz}$, ${xy}$, and $z^2$ states are visible using $p$-polarized light as used in Fig.~\ref{1}.
From the DFT calculations reported in detail later in the paper, it transpired that the third, $\Gamma$-centered hole pocket, $\alpha_3$, that we were unable to distinguish in the present data has mainly $xy$ orbital character.
Therefore we assign the bands $\alpha_1$ and $\alpha_2$, detected using $p$-polarized light to have mainly $xz$ and $z^2$ orbital character.

Figure~\ref{2x} depicts analogous data to Fig.~\ref{1} but now recorded using $s$-polarized light. In Fig.~\ref{2x} we could again resolve two bands at the zone center: $\alpha_1$ and $\alpha_2$.
As in the data shown in Fig. 1, the $\alpha_1$ disperses strongly towards $E_F$, and the $\alpha_2$ feature crosses the Fermi level at a momentum vector of $k_F$=0.15$\pm$0.02~$\AA^{-1}$.
In contrast to the data shown in Fig.~\ref{1}, we did not observe any spectral weight at the $Z$-point for the $s$-polarized case.  
In this measurement geometry, $s$-polarized light would be expected to detect bands having ${x^2-y^2}$ and ${yz}$ orbital characters. As we know that the ${x^2-y^2}$ states are located far below the Fermi level at the zone center,~\cite{Thirupathaiah2010} we exclude these states from further discussion. Hence, the bands $\alpha_1$ and $\alpha_2$ shown in Fig.~\ref{2x} have predominantly only $yz$ orbital character.

From  Figs.~\ref{1} and ~\ref{2x} it is clear that the spectral intensity of the hole pocket $\alpha_2$ at the zone center is elongated in the $k_{y}$ direction when probed with $p$-polarized light and is elongated in the $k_{x}$ direction when measured using $s$-polarized light. This observation suggests that the orbital contribution to the $\alpha_2$ Fermi sheet is directional, i.e., in the $k_y$ direction the FS sheet has predominantly $xz$ orbital character and in the $k_x$ direction it is predominantly of $yz$ character. This observation is in very good agreement with the predictions made in Ref.~\onlinecite{Graser2009}. Note here that the orbital contribution to the $\alpha_1$ Fermi sheet will be the other way round, meaning that in the $k_y$ direction this FS sheet has predominantly $yz$ character and in the $k_x$ direction it is predominantly composed by the $xz$ character, as reported in Ref.~\onlinecite{Graser2009} . The directional orbital contribution to this Fermi surface is predicted by theory for the iron pnictide compounds, but in experimental data, the presence of an $xy$ hole pocket with circular energy contours does cast some doubt on this, when viewed from the perspective of the ARPES data of the iron pnictide system.

In Figs.~\ref{1} (d) and (f), a broad spectral feature labelled $\gamma$ can be seen at a binding energy E$_B$=0.35 eV that is not seen when the experiment is conducted with $s$-polarized light.
A very similar band dispersion has been observed experimentally in BaFe$_2$As$_2$,~\cite{Fink2009} but at the greater binding energy E$_B$=0.6 eV, and is ascribed to the band formed by the ${z^2}$ states, thus we follow this 
attribution here also for the Fe chalcogenide.~\cite{Fink2009}

Within this picture, the band $\gamma$ is shifted almost 250 meV towards the Fermi level compared to Ba122,~\cite{Fink2009}, indicating a different hybridization between the Fe 3$d$ states and the chalcogenide 4$p$ states in these compounds compared to the 122 materials.~\cite{Turner2009} This conclusion is also consistent with the earlier ARPES data on stoichiometric and non-stoichiometric Fe chalcogenide and related compounds, as well with DFT calculations.~\cite{Singh2010a, Subedi2008c,Chen2010a,Nakayama2010}

\begin{figure}[b]
	\centering
		\includegraphics[width=0.48\textwidth]{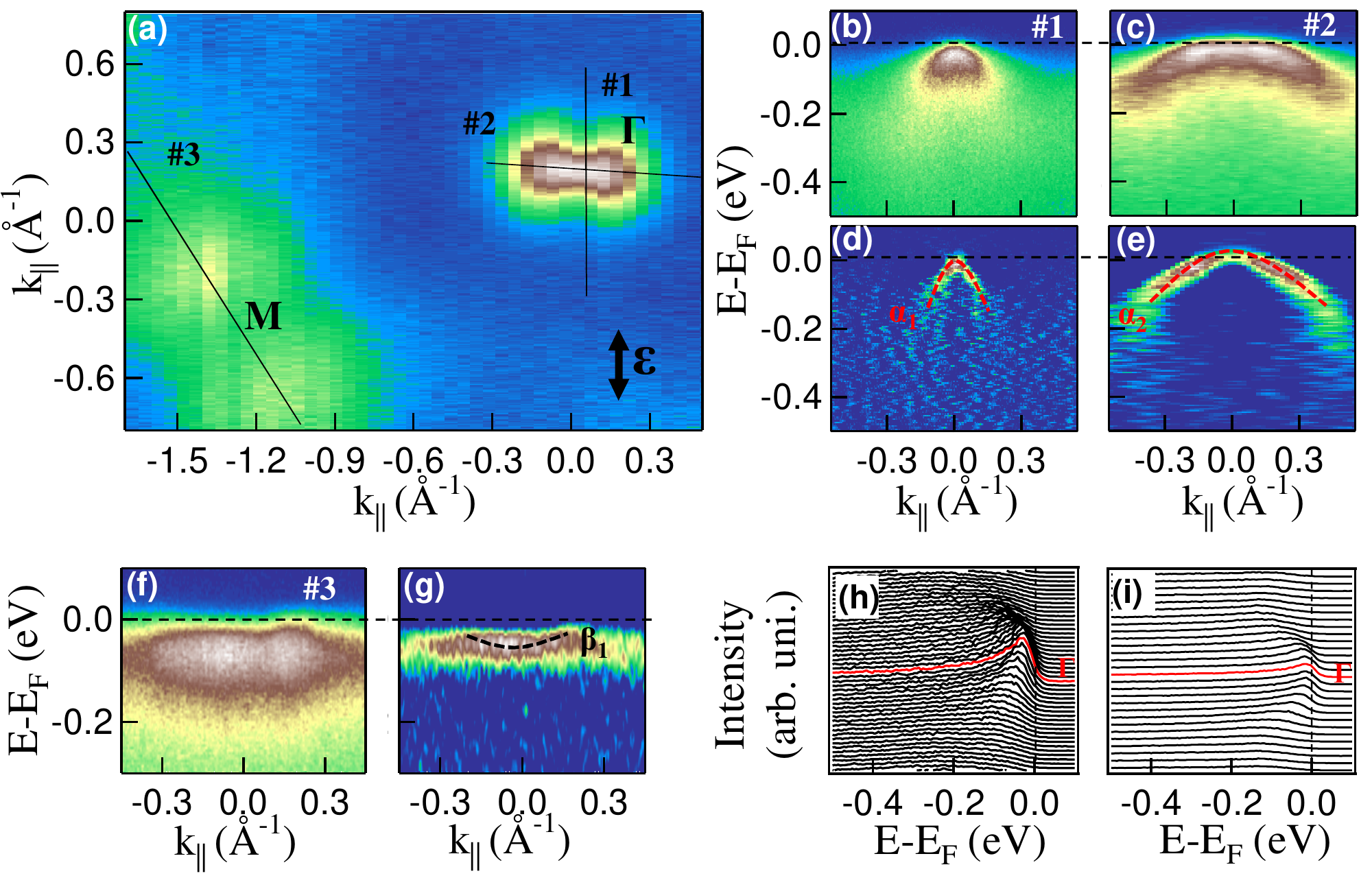}
	\caption{ARPES data from Fe$_{1.068}$Te$_{0.64}$Se$_{0.36}$ measured with an excitation energy h$\nu$=88 eV using $s$-polarized light. Panel (a) shows the Fermi surface map. On the figure, the light polarization vector ($\vec{\varepsilon}$) is displayed. Panels (b), (c) and (f) show the energy distribution maps (EDMs) taken from the cuts \#1, \#2 and \#3, respectively, as shown overlaid on the FS map. Panels (d), (e) and (g) are the second derivatives of (b), (c) and (f), respectively. (h) and (i) show energy dispersive curves from EDMs in (b) and (c), respectively.}
	\label{3}
\end{figure}

Figure~\ref{3} shows ARPES data from Fe$_{1.068}$Te$_{0.64}$Se$_{0.36}$ measured with an excitation energy of h$\nu$=88 eV using $s$-polarized light. In  Fig.~\ref{3}(a) we show the FS map extracted from integrating over an energy window of 10 meV centred at $E_F$, in which hole pockets at the zone center and an electron pocket at the zone corner are seen, similar to the data from the crystals with x=0.46. Figs.~\ref{3}(b) and ~\ref{3}(c) show EDMs taken along the cuts $\#$1 and $\#$2, respectively. From these EDMs, two hole-like bands, $\alpha_1$ and $\alpha_2$, can be resolved at the zone center. The band $\alpha_1$ disperses strongly towards $E_F$ but does not cross it, while $\alpha_2$ crosses $E_F$ at a Fermi vector ($k_F$) 0.15$\pm$0.02~$\AA^{-1}$. Fig.~\ref{3}(f) shows the EDM resulting from cut $\#$3 in which an electron-like band we label $\beta_1$ can be seen at the zone corner. The second derivative of the EDM from panel (f) is shown in Fig.~\ref{3}(g). Both the raw data and the second derivative show that the bottom of the electron pocket is close to $E_F$, indicating that the electron pocket is shallow, as has also seen in other iron-based superconductors.~\cite{Thirupathaiah2013,Maletz2014,Starowicz2013}

\begin{figure}[b]
	\centering
		\includegraphics[width=0.4\textwidth]{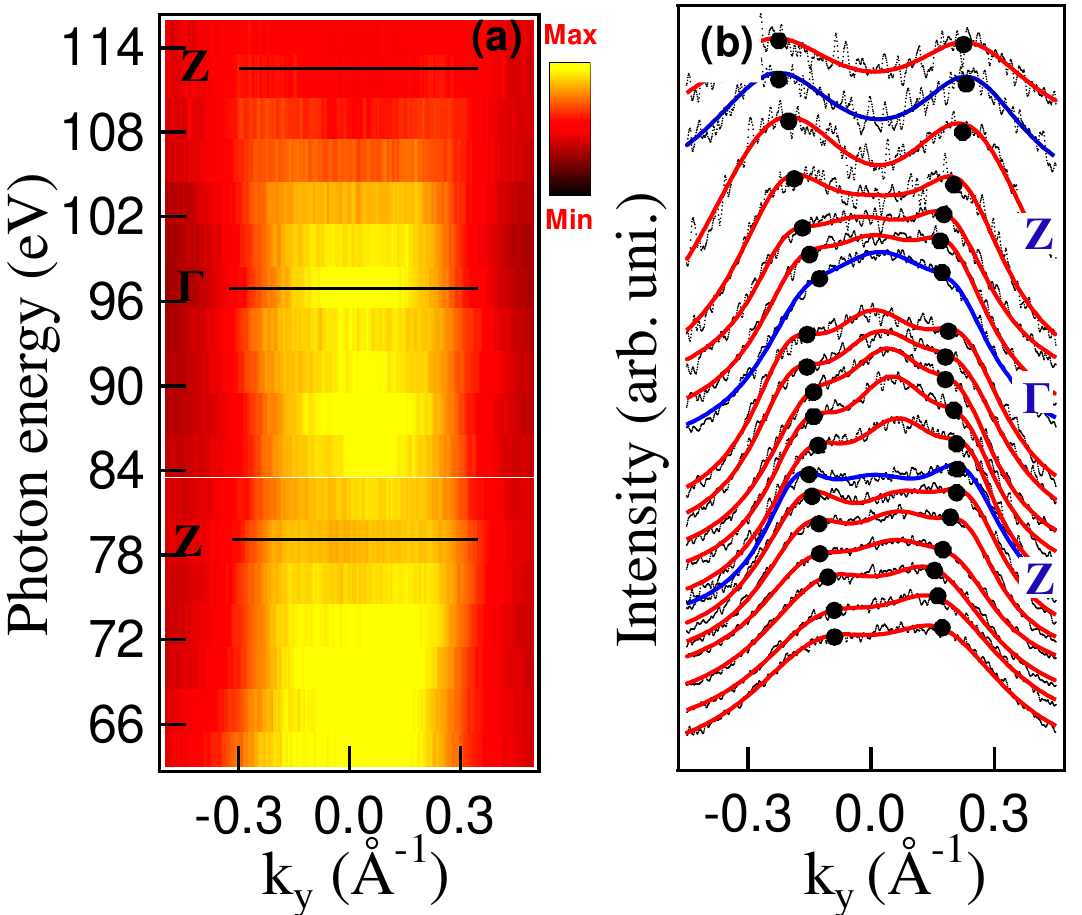}
		\caption{(Color online) Photon energy dependent data taken from Fe$_{1.068}$Te$_{0.54}$Se$_{0.46}$ to reveal the $k_z$ dependence of the electronic structure at the zone center measured using $p$-polarized light. Panel (a) shows the $k_y$, $k_z$ Fermi surface map extracted over an integration window of 10 meV centred at $E_F$, with the high symmetry points in $k_z$ marked. Panel (b) shows a stack-plot of momentum dispersion curves (MDCs) sampling different $k_z$, together with the results of a fit using two Lorentzian functions near the $Z$-point and using three Lorentzian functions near the $\Gamma$-point. The black circles overlaid on the MDCs in panel (b) represent the peak positions of the $\alpha_2$ band.}
	\label{4}
\end{figure}

Next we show ARPES measurements performed to reveal information on the $k_z$ dependent electronic structure. For this, photon energy dependent ARPES spectra were recorded for $k_{||}$ near the zone center, with photon energies ranging from h$\nu$=63 to 117 eV in steps of 3 eV. Data were recorded using  $p$-polarized light along the $\Gamma$-$X$ high symmetry line. Fig.~\ref{4}(a) depicts the Fermi surface map in the $k_y-k_z$ plane.  Figure~\ref{4}(b) shows momentum distribution curves as a function of photon energy, fitted with two or three Lorentzian functions. The peak positions of the $\alpha_2$ band extracted from the fits are shown by the black cirlces on the MDCs. The high symmetry points $\Gamma$ (h$\nu$ = 96 eV) and $Z$ (h$\nu$ = 81 and 114 eV) have been  identified using the formula

\begin{eqnarray}
	k_{\bot} = \sqrt{\frac{2m_e}{\hbar ^2} [E_{kin} cos^2\theta+V_0]}\hspace{1 mm}
	\label{eq1},
\end{eqnarray}
where the inner potential, $V_0$, has been taken to be 15$\pm$2 eV.~\cite{Thirupathaiah2010}

Figs.~\ref{5} (a)-(c) show the EDMs taken along the $\Gamma$-$X$ high symmetry line at $k_z$ = 0, 0.5 and 1 in units of $\pi/c$, where $c$ is the c-axis lattice parameter. These data were measured using h$\nu$ = 96, 90 and 81 eV, respectively, and with $p$-polarized light. Superimposed in black on panels ~\ref{5} (a)-(c) are the dispersion relations of the hole-like band, $\alpha_2$, estimated from the fit to the MDC curves using two Lorentzian functions. The white dashed lines represent a parabolic fit to the black, MDC-derived curve.
Similarly, Figs.~\ref{5} (d)-(f) show analogous EDMs recorded at $k_z$ = 0, 0.5 and 1 ($\pi/c$), measured using the photon energies h$\nu$ = 96, 90 and 81 eV, with $s$-polarized light. Superimposed in black on panels ~\ref{5} (d)-(f) are the dispersion relations of the hole-like band, $\alpha_1$, estimated from the fit to the MDC curves using two Lorentzian functions. The white dashed lines again represent a parabolic fit to the black, MDC-derived curve. 
Figs.~\ref{5} (g)-(i) show the hole-like band dispersions from the DFT band structure calculations along the $\Gamma - X$ high symmetry direction in $k_{||}$ for $k_z$ = 0, 0.5 and 1 ($\pi/c$), respectively. 
In panels (g)-(i) the dashed-curves are hole-like bands from the calculations, while the red/blue solid lines are the results of the parabolic fit to the experimental bands corresponding to $\alpha_1$/$\alpha_2$. The Fermi level of the calculated bands is shifted such that the Fermi wavevector of the $\alpha_2$ hole pocket matches that seen in experiment. In this way it is easier to calculate the renormalization of the bands. However, this method may lead to discrepancy in estimating the renormalization of the $\alpha_1$ band (which does not cross $E_F$). We will discuss this point in detail in the next section.
\\
\subsection{Spectral functions and self-energies}
\subsubsection{Theory}

ARPES provides an experimental window on the single particle spectral function, $A(E,k)$, and with a complex self-energy $\Sigma(E,k)=\Re\Sigma(E,k)+i\Im\Sigma(E,k)$ it is given by

\begin{eqnarray}
A(E,k) = -\dfrac{1}{\pi} \dfrac{\Im\Sigma}{(E_k-\epsilon(k)-\Re\Sigma)^2+(\Im\Sigma)^2},
\label{eq2}
\end{eqnarray}

where the real part of self-energy $\Re\Sigma(E,k)$ can be extracted by subtracting the bare-band dispersion $\epsilon(k)$ from the experimentally determined, renormalized band dispersion ($E_k$): $\Re\Sigma(E,k)=E_k-\epsilon(k)$.
The imaginary part of the self-energy can be extracted from the momentum widths of the experimental band features $\Delta_k$ and the bare-band velocity $v_k$, is given by
\begin{eqnarray}
\Im\Sigma(E,k)=\Delta_k v_k,
\label{eq2x}
\end{eqnarray}

in which $\Delta_k$ is the half-width half maximum of the momentum distribution curve.
On the other hand, one can also calculate the imaginary part of the self-energy using the scattering rate $S(E)=\Delta_k v^*_k$ and mass renormalization ($m^*/m_b$) using the expression 

\begin{eqnarray}
\Im\Sigma(E) = S(E)\dfrac{m^*}{m_b}.
\label{eq3}
\end{eqnarray}

Here $v^*_k$ is the renormalized velocity and it is assumed that $\Im\Sigma(E)$ depends only weakly on the momentum $k$. $m^*$ is the effective mass estimated from the experimental band structure and $m_b$ is the bare-band mass estimated from the calculated band structure.

There are several theoretical approaches to describe non-Fermi liquid behaviour of the single-particle spectral function that can be observed in ARPES. For example, using purely phenomenological ansatz, marginal Fermi liquid theory (MFL)~\cite{Varma1989} gives:

 \begin{eqnarray}
 \Sigma(E)^{MFL} = \dfrac{1}{2}[\lambda_{MFL}E\ln(\dfrac{E_c}{u})-i\pi\lambda_{MFL}u],
 \label{eq4}
\end{eqnarray}

which is often used in fitting the ARPES data of high-$T_c$ cuprates.~\cite{Chang2013,Damascelli2003} Here $u = max(|E|, k_BT)$, where $k_BT$ is the thermal energy. $E_c$ is the cutoff energy, which in a first approximation corresponds to the width of the conduction band.~\cite{Varma1989} Note here that in context of the marginal-Fermi liquid theory, the scattering rate can be expressed as $S(E)=\alpha+\beta E$, where $\alpha$ represent the elastic electron-impurity scattering processes and $\beta$ represents the electron-electron inelastic scattering.
On comparing Eqs.~\ref{eq2} and ~\ref{eq4}, and considering the linear dependence of $S(E)$ on the energy,  we can then calculate the electron coupling constant using the formula

 \begin{eqnarray}
 \lambda_{MFL} = \dfrac{2}{\pi}\dfrac{m^*}{m_b}\beta.
 \label{eq5}
 \end{eqnarray}

This marginal-Fermi liquid behaviour naturally emerges in microscopic theories near the quantum critical point in 3D systems. However obtaining this behavior in 2D systems remains problematic.

Another scenario for non-Fermi liquid behaviour is based on the idea that the dominant interaction in the cuprates is between the fermions and their low-energy collective spin excitations. In this scenario, the non-Fermi liquid behavior in the normal state is associated with the proximity to a critical point, but this point now separates paramagnetic and antiferromagnetically ordered phases.
It has been shown in the past~\cite{Haslinger2002} that in this case the self-energy can be written as

\begin{equation}
\Sigma^{sf}(E)=\lambda_{sf}\frac{2E}{1+\sqrt{1-i|\frac{E}{\omega_{sf}}|}}
\label{eq6}
\end{equation}

At small energies, $E<<\omega_{sf}$, the system displays Fermi-liquid behavior but is non-Fermi-liquid-like for intermediate and frequencies well above $\omega_{sf}$

\subsubsection{Application of the theory to the ARPES data}

In Fig.~\ref{6} we show the spectral width analysis of the data measured on the Fe$_{1+y}$Te$_{0.5}$Se$_{0.5}$ sample. From Fig.~\ref{6} (c) it is clear that energy dependent scattering rate obtained near the zone center suggests a non-Fermi liquid behaviour for the quasiparticles populating the $\alpha_1$ band,  specifically a marginal-Fermi liquid type behaviour.  As an example, if Eq.~\ref{eq6} (spin-fluctuation, SF) is applied to extract the self-energy of the hole pocket near the $\Gamma$ point, we find that the expression is able to give a very good agreement to the data, as shown in Fig.\ref{7}. On the other hand,  the qualifier is that for the spin-fluctuation theory, an unrealistically large value of $\lambda_{sf} \sim 7$ with $\omega_{sf}=30~meV$ is required to get this good fit. We note, however, that the SF-theory expression used here does refer to the single band case, while in the multiband situation relevant for Fe$_{1+y}$Te$_{1-x}$Se$_x$, the quasiparticle linewidth is determined by the sum of intraband and interband interactions and therefore the absolute numbers for $\lambda_{sf}$ inferred from the single-band theory should be taken with caution. In addition, spin-fluctuation theory predicts Fermi-liquid behaviour at energies well below $\lambda_{sf}$ (here $\approx$ 15 meV), a behavior that is not resolved in these data at present. Due to these facts, we chose in the following to concentrate on the MFL expression ( Eq.~\ref{eq5}) for the analysis of data, without specifying the microscopic origin of its self-energy. Given the MFL picture, the electron coupling constant $\lambda_{MFL}$ extracted is 1.5 near the zone center. 
This value matches well with the MFL coupling constant, $\lambda_{MFL}$=1.6, extracted from ARPES data recorded from doped BaFe$_2$As$_2$ and NaFeAs iron pnictides.~\cite{Fink2015} 
	
\begin{figure}
	\centering
		\includegraphics[width=0.49\textwidth]{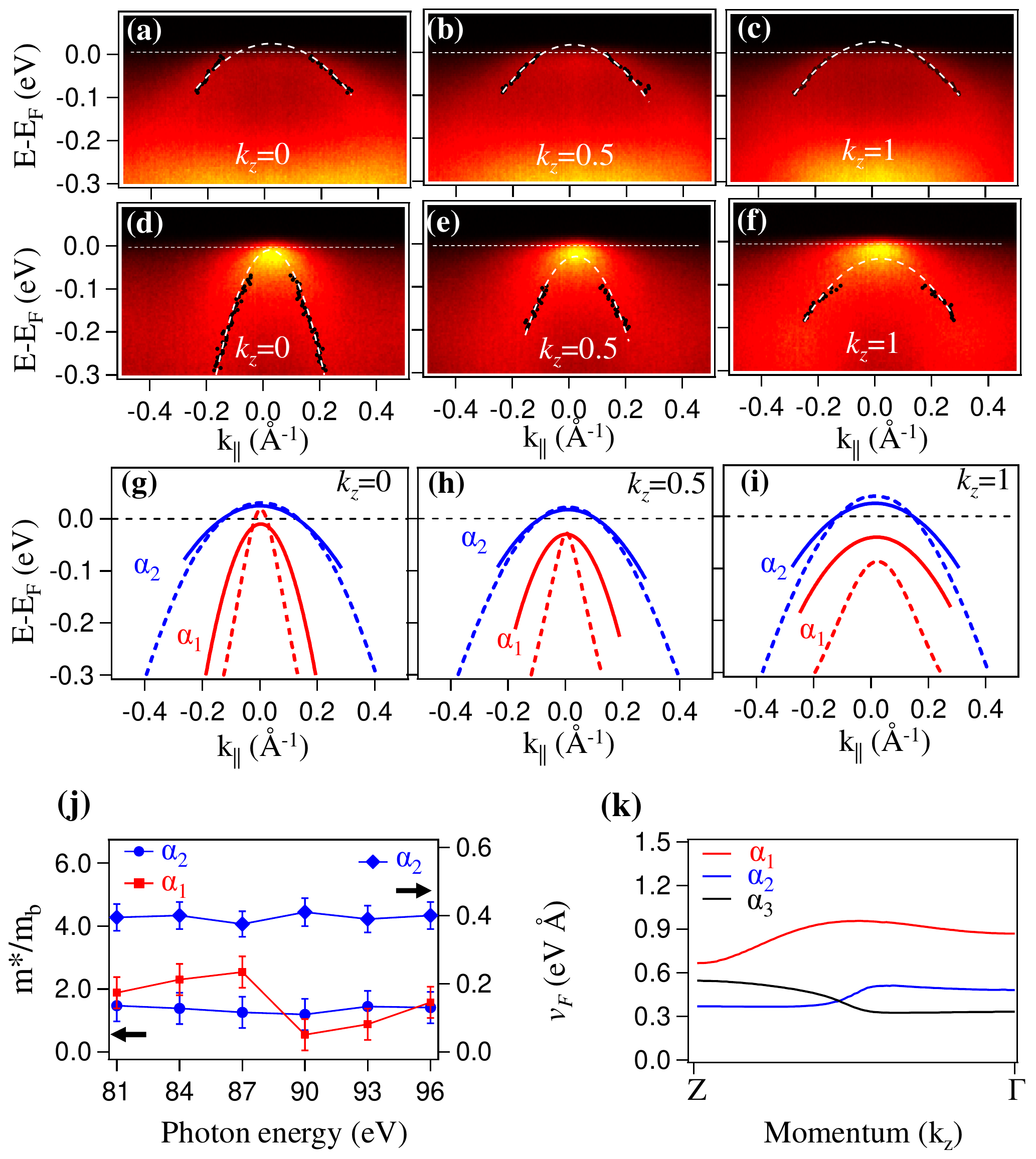}
		\caption{(Color online) ARPES data taken on Fe$_{1.068}$Te$_{0.54}$Se$_{0.46}$. EDMs shown in (a)-(c) are measured at the $k_z$ values indicated (in units of $\pi/c$), using $p$-polarized light, and show the dispersive $\alpha_2$ band. Panels (d)-(f) show analogous EDMs, but measured using $s$-polarized light, and show the $\alpha_1$ band. In all panels (a)-(f), the black dotted curves result from a fit to the MDC's using a pair of Lorentzian functions, and the thin white dashed curves shows parabolae fitted to the black dotted dispersion curves. From these parabolae, the effective mass, $m^*$, can be determined experimentally. Panels (g)-(i) show the results of DFT band structure calculations performed on the parent FeSe compound,~\cite{Eschrig2009} and the dashed lines show the pair of hole-like bands predicted for each $k_z$ value. The parabolic fits to the experimental band dispersions corresponding to $\alpha_1$ and $\alpha_2$ are shown in panels (g)-(i) as red and blue solid lines, respectively. Panel (j) depicts the $k_z$ dependence (probed via changing the photon energy) of the mass renormalization ($m^*/m_b$) for the $\alpha_1$ (red) and $\alpha_2$ (blue) bands and Fermi velocity (upper curve, $v_F$) for the $\alpha_2$ band. Panel (k) shows the calculated $k_z$ dependence of the Fermi velocity ($v_F$) for the three hole-like bands estimated from the DFT calculations.}
	\label{5}
\end{figure}

\begin{figure}
	\centering
		\includegraphics[width=0.49\textwidth]{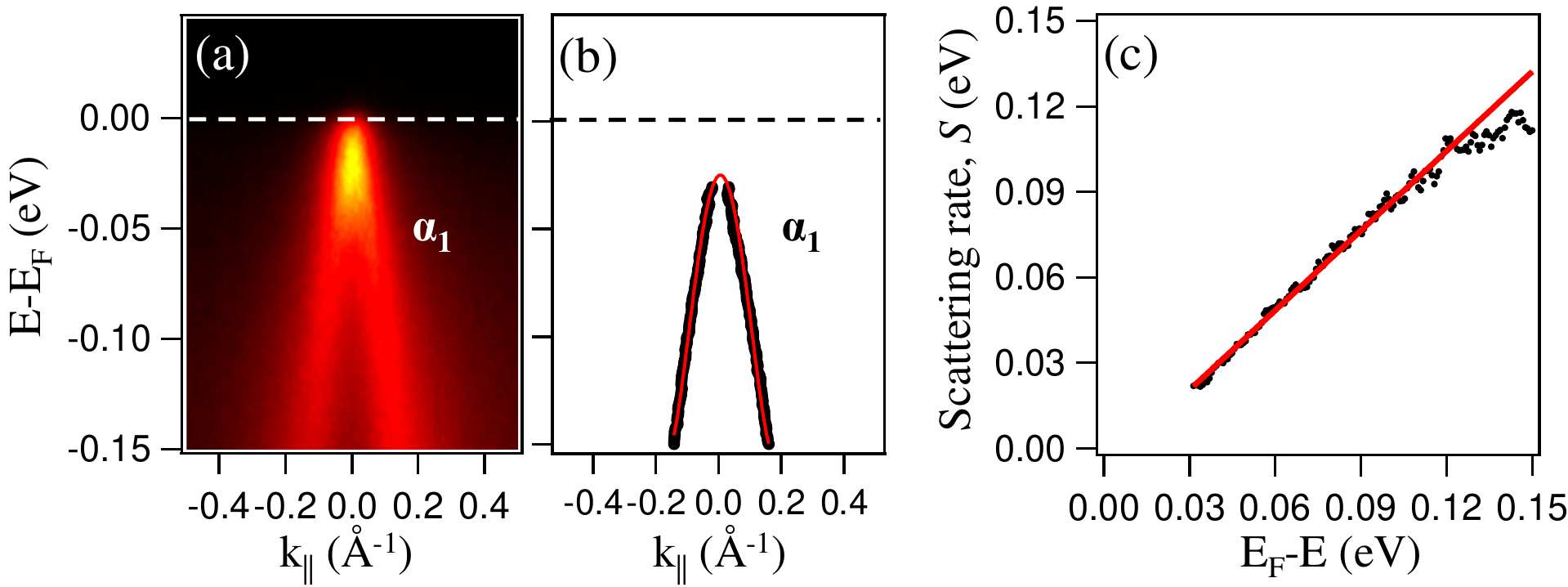}
		\caption{(Color online) EDM shown in panel (a) is taken from the Fe$_{1+y}$Te$_{0.5}$Se$_{0.5}$ sample measured using $s$- polarized light with a photon energy of 46 eV, which corresponds to $k_z$=0. Black curves in panel (b) are the experimental band dispersions extracted from fitting Lorentzian functions to the momentum dispersive curves from the data shown in panel (a). The red curve shows a fit to the experimental hole-like band using a 4$^{th}$ order $E-k$ dispersion relation.
		The black curve in panel (c) is the energy dependent scattering rate extracted from the data shown in (a) and the red curve shows the result of a fit using marginal-Fermi-liquid theory.}
	\label{6}
\end{figure}

\section{Discussion}
\subsection{Fe non-stoichiometry}
Three well-resolved hole pockets around $\Gamma$ have been reported in ARPES data from an iron-stoichiometric FeTe$_{0.42}$Se$_{0.58}$~\cite{Tamai2010, Yi2015,Lubashevsky2012} superconductor, while the data presented here from our non-Fe-stoichiometric Fe$_{1.068}$Te$_{1-x}$Se$_{x}$ ($x$=0.36 and 0.46) superconductors only contain two hole pockets at the zone center.
This difference matches with other published data on non-Fe-stoichiometric Fe$_{1.03}$Te$_{0.7}$Se$_{0.3}$~\cite{Nakayama2010} and Fe$_{1.03}$Te$_{0.94}$Se$_{0.6}$~\cite{Lubashevsky2012} compounds, in which only two hole pockets were observed in ARPES at the zone center.
We did pick up three hole pockets from data (not shown)  measured on  the close to Fe-stoichiometric Fe$_{1+y}$Te$_{0.5}$Se$_{0.5}$ (y \textless 1$\%$) sample, a result consistent with data from the Fe-stoichiometric FeTe$_{0.56}$Se$_{0.44}$ compound.~\cite{Yi2015}. 

One recent ARPES report on the stoichiometric FeTe$_{0.56}$Se$_{0.44}$ suggested that upon increasing the sample temperature the hole pocket with $xy$ character completely loses its spectral weight, while the other two pockets ($xz/yz$ and $z^2$) maintain their itinerant character also at higher temperature.~\cite{Yi2015} This was explained as the evolution to an orbitally-selective Mott-insulator at higher temperature. The data presented here are measured at a sample temperature close to 1K, and yet the third hole pocket around $\Gamma$ is already missing in the case of Fe$_{1.068}$Te$_{1-x}$Se$_{x}$ compounds.

The FeSe and FeTe systems and their doped variants display complex and rich defect chemistry. For example, in Ref.~\onlinecite{Chen2014}, ordering of Fe vacancies in $\beta$-Fe$_{1-x}$Se is argued to lead to a non-superconducting, 'parent' phase of the FeSe superconductors. In Ref.~\onlinecite{Wang2015}, K$_2$Fe$_4$Se$_5$ is argued to be an Fe vacancy-ordered non-superconducting parent compound to the high-T$_c$ K-intercalated FeSe superconductors.  
Thus, the issue of off-stoichiometry in these systems is central to their electronic structure and ground-state properties.  

Comparing the electronic structure between stoichiometric and  non-stoichiometric compounds it can be seen that already an iron excess of only 3$\%$ - irrespective of the amount of Se doping - is enough to lead to the third hole pocket at the zone center being barely resolvable.~\cite{Tamai2010,Nakayama2010, Yi2015, Chen2010a} 
The absence of the third hole pocket (that one which has dominant $xy$ character) could be linked to its Mott-insulating character due to the interaction with the local magnetic moment of the excess iron. As the because of which the spectral weight of $xy$ band is totally lost compared to the $xz/yz$ bands,~\cite{Zhang2010f} the interaction between the itinerant electrons and the local magnetic moment of excess iron would seem to have more effect on the in-plane $xy$ band compared to the $xz/yz$ bands which possess more out of plane character. 
A theoretical study suggested that each excess iron atom provides an additional electron to the system in these compounds,~\cite{Zhang2009e} which could be expected to give rise to a rigid-band-type shift of the Fermi level. This kind of behaviour has been seen on electron doping in the 122 iron pnictide systems.~\cite{Thirupathaiah2010}
From a comparison of our 11 ARPES data with those of Ref.~\onlinecite{Nakayama2010} and Ref.~\onlinecite{Tamai2010}, we notice that the $\gamma$ band [as seen in Fig.~\ref{1} (d-g)] has a constant binding energy of 0.35 eV, irrespective of the amount of the excess Fe present in the composition. This would argue against a simple rigid-band-type scenario for the excess iron in the 11 compounds.

\subsection{Isovalent Se,Te substitution}
Isovalent substitution generally induces an additional crystal field potential to the system, and therefore, could lead to changes in the electronic structure as has been seen in the iron pnictide 122 system (BaFe$_2$As$_2$) on P substitution for As~\cite{Thirupathaiah2011} or Ru for Fe.~\cite{Brouet2010} 
In our present study, Se substitution at the Te site is also isovalent doping that could lead to a crystal field splitting of the Fe $3d$ orbitals. Hence, one may expect changes in the electronic structure of Fe$_{1.068}$Te$_{1-x}$Se$_{x}$ with varying Se doping concentration.
However, we did not observe noticeable changes for $x$ varying between $x$=0.36 and 0.46 (see Figs.~\ref{1},~\ref{2x} and ~\ref{3}).
In the case of the iron pnictides, we have seen that the isovalent substitution of P for As in the Ba122 system leads to changes in the electronic structure even for a substitution as small as 5$\%$.~\cite{Thirupathaiah2011}
A recent report on the iron chalcogenides offers a solution to this apparent discrepancy, as it communicates that Se doping mainly affects the band of $xy$ character, leaving the other two hole-like bands ($xz/yz$ and $z^2$ at the zone center) mostly unchanged.~\cite{Liu2015}
As already discussed above, our ARPES data show only two hole-like bands around the zone center, and our polarisation analysis attributes these to the $xz/yz$ and $z^2$ related bands, so the arguments of Ref. \onlinecite{Liu2015} also fit our data well.  

\subsection{Orbital ordering}
A directional orbital contribution to the hole pockets in the iron pnictides has been proposed by Graser $et~al$., in their itinerant picture of the electronic structure of these systems.~\cite{Graser2009} 
In the present study of the 11 system, a directional orbital contribution to the hole pockets could clearly be observed [see Fig.~\ref {1} and Fig.~\ref {2x}]. What the implications are of this orbital ordering in $k$-space for superconductivity is not clear at present. Intraorbital interactions between hole and electron pockets have been argued to be advantageous for iron-based superconductivity, over interorbital interactions.~\cite{Saito2010} In this context the orbital ordering of the Fermi sheets that contribute to the Cooper pairs at both the center and corner of the Brillouin zone is certainly an asset for high-$T_c$ superconductivity, in addition to considerations involving Fermi surface nesting.

\begin{figure}[b]
	\centering
		\includegraphics[width=0.48\textwidth]{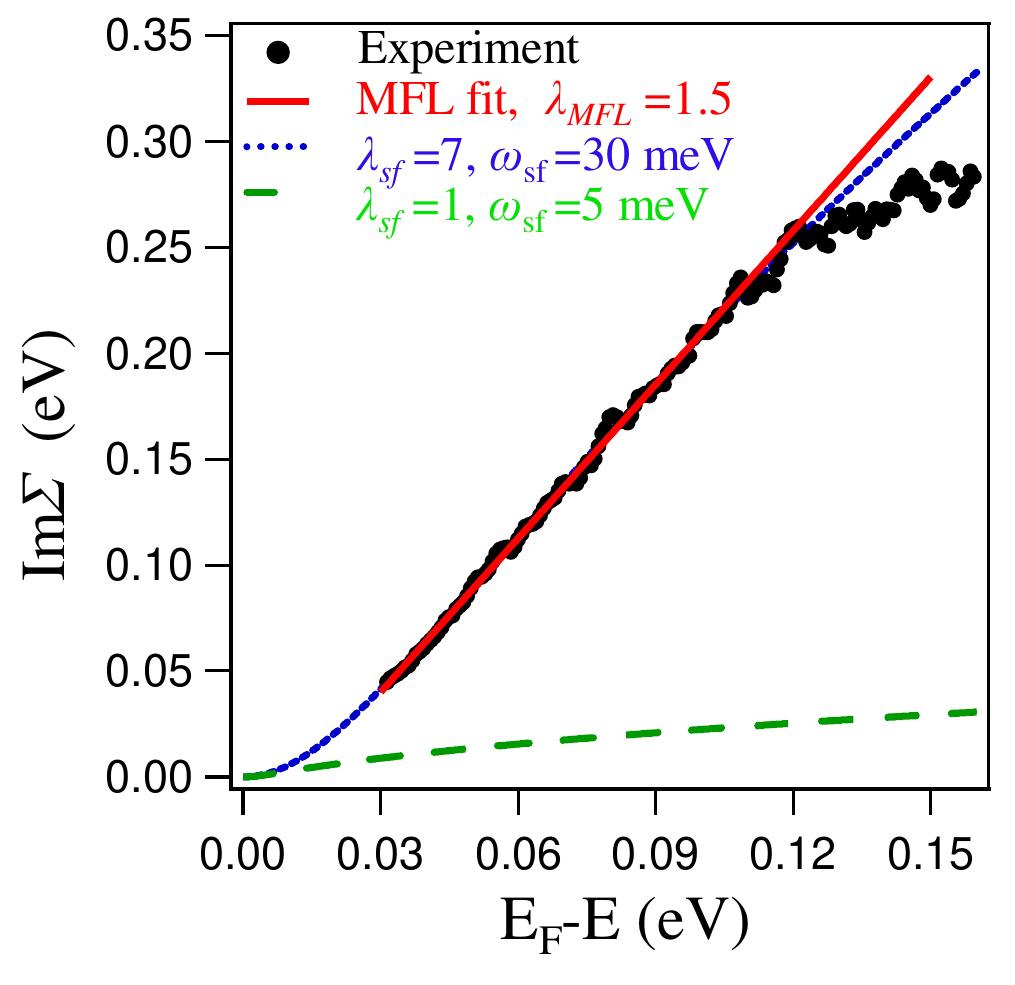}
		\caption{(Color online) Imaginary part of self-energy ($\Im\Sigma$) is plotted as a function of energy below $E_F$. The different lines compare $\Im\Sigma$ from marginal-Fermi-liquid theory and spin-fluctuation theory. Black curve is the experimental data and red solid curve is a MFL fit to the experimental data shown in Fig.~\ref{6}(c).  Blue and green dashed curves are the SF-theory simulations using Eq.~\ref{eq6}, and differ with respect to their coupling constant ($\lambda_{sf}$) and characteristic energy ($\omega_{sf}$).}
	\label{7}
\end{figure}

\subsection{Dependence of the electronic structure on $k_z$}
\subsubsection{Orbital character}

Next, we discuss the $k_z$ dependence of the electronic structure at the zone center. In 122 systems, we earlier reported a strong $k_z$ Fermi surface warping at the zone center due to the transformation of orbital character from $xz/yz$ to $z^2$ while going from $\Gamma$ to $Z$.~\cite{Fink2009} In contrast, Fe$_{1.068}$Te$_{0.54}$Se$_{0.46}$ shows only a weak $k_z$ warping along the $\Gamma-Z$ direction (see Fig.~\ref{4}), a situation also picked up on in Ref.~\onlinecite{Starowicz2013}. This can be linked to the absence of an orbital character switch from $xz/yz$ to $z^2$ in the 11 compounds, as opposed to the 122 systems. This conclusion is supported by the observation of $k_z$ dependent band dispersion of the $\alpha_1$ band [see Figs.~\ref{5} (d)-(f)], attributed here with the help of the DFT calculations to the $xz/yz$ and $z^2$ orbital character.

From Figs.~\ref{5} (d)-(f) it can clearly be seen that the $\alpha_1$ band just touches $E_F$ at the $\Gamma$-point and then disperses away from the Fermi level towards higher binding energy while approaching the $Z$-point. Therefore, the $z^2$ orbital does not contribute to the Fermi surface at the $Z$ point, meaning that the states seen in the $k_z$ map measured near the zone center have solely $xz/yz$ character [see Fig.~\ref{4} (a)]. 
These observations are in good agreement with the minimal orbital theory of iron-based superconductors,~\cite{Daghofer2010a} which stresses not only the simple crystal structure of the iron chalcogenide superconductors but also their simple low-energy electronic structure.
Two further interesting points can be noted here: (a) in this 11-compound, only a single band exists at the Fermi surface at the $Z$-point which could contribute to superconductivity, whereas in 122 systems all three bands are present and (b) the top of the $\alpha_1$ band shifts towards higher binding energy in the present system while going from $Z$ to $\Gamma$, whereas it shifts towards lower binding energy in the 122 systems while going from $Z$ to $\Gamma$.

\subsubsection{Mass renormalization}
From the estimation of mass renormalization as a function of photon energy shown in Fig.~\ref{5}(j), it can be seen that the $\alpha_2$ band retains a value of $m^*/m_b$ $\approx$ 1.8$\pm$0.3 for all $k_z$ values probed. 
This is in contrast to the case for the $\alpha_1$ band, which shows strong variation in the mass renormalization from $m^*/m_b$ $\approx$ 1.5$\pm$0.4 to 5.2$\pm$1 in the region for which $k_z$=0.5 (h$\nu$ = 87 and 90 eV).
We note that a $m^*/m_b$ value of just under two is shared by both bands close to $\Gamma$ (h$\nu$ = 96 eV).
As mentioned previously in the results section, the $\alpha_2$ band from DFT was shifted so as to match the experimental $k_F$ for this band. This could not be done for the $\alpha_1$ band, and the resultant uncertainty in the fidelity of the energy location of the top of this band in the DFT could contribute to the observed strong variation in the mass renormalization for $\alpha_1$.

There is good consistency between the mass renormalization and the calculated Fermi velocity for the band $\alpha_1$. Fig.~\ref{5}(k), which changes in Fermi velocity from 0.65 eV\AA  ~at the $Z$-point to greater than 0.9 mid-way to $\Gamma$ and finally takes a value of 0.85 eV\AA  ~at the $\Gamma$-point itself.
In contrast, the DFT predicts a $k_z$ independent Fermi velocity of $v_F$=0.5$\pm$0.1 eV\AA ~for the $\alpha_2$ band, and this is not only quantitatively consistent with the experimental data that give a $k_z$-independent $v_F$=0.4$\pm$0.1  eV\AA ~[see Fig.~\ref{5}(j)], but also consistent with the $k_z$ independent mass renormalization for this band.
We note here that, on the whole, the mass renormalizations we observe for both hole pockets are consistent with the values of $m^*/m_b$ $\approx$ 2-4 reported in Refs.\onlinecite{Xia2009a,Liu2015}. 
Closing the discussion on the effective mass, we emphasize that in the light of the calculations reported in Ref.~\onlinecite{Fink2015}, the moderate mass enhancements seen here of between 2 and 4 occur only at higher binding energies, i.e. well away from the chemical potential.
In the case where a flat band lies close to the Fermi level yielding a van Hove singularity and there is an imminent Lifshitz transition, then a dramatic increase in the mass enhancement occurs within the marginal-Fermi liquid model, which directly follows from the linear-in-energy dependence of imaginary part of the self-energy ($\Im\Sigma$). This means that when calculating the real part of self-energy ($\Re\Sigma$) via a Kramers Kronig transformation of $\Im\Sigma$, the low-energy logarithmic increase of $\Re\Sigma$ leads to a very flat band and to strong mass enhancements of order 10 near the chemical potential.~\cite{Fink2015}

\subsubsection{Quantum criticality and energy dependent scattering rates}
Quantum criticality in the iron-based superconductors is part of the current consensus as regards the understanding of high-$T_c$ superconductivity in these materials. A quantum critical point in these compounds has been observed experimentally~\cite{Kasahara2010,Meingast2012, Ning2010a,Shishido2010} and predicted theoretically.~\cite{Dai2009a, Abrahams2011} Quantum criticality in iron-based superconductors is rooted to short range spin-fluctuations active across an interband nesting vector ($\pi$,0). Near the quantum critical point, the system switches from being a Fermi liquid to displaying marginal Fermi liquid behavior. This means that the imaginary part of the self energy has a linear dependence on the energy,~\cite{Varma1989, Johnson2007} which is significantly different from the quadratic energy dependence observed in conventional Fermi liquids. In the present case this has been systematically studied for the 11 system.
Earlier DMFT calculations suggested a crossover from Fermi liquid to a non-Fermi liquid character in the case of BaFe$_2$As$_2$ at optimal hole doping given sufficiently high sample temperatures.\cite{Werner2012} No such behaviour has been predicted with temperature for electron doping in the 122 materials, and recent optics data show Fermi liquid behavior in the bulk of annealed, electron doped Ba122 crystals.~\cite{Tytarenko2015}

On the contrary, a recent ARPES study on various 122 and 111 systems doped with charge carriers and with isovalent substitution into the parent compound unambiguously shows a non-Fermi-liquid character near a regime of optimal charge doping or substitution.~\cite{Fink2015} 
In the ARPES data presented here from the iron chalcogenide Fe$_{1+y}$Te$_{0.5}$Se$_{0.5}$ system, a non-Fermi-liquid behaviour of the quasiparticles was found for the band $\alpha_1$ near the zone center by extraction of the scattering rates as a function of the binding energy [see Fig.~\ref{6}]. Specifically, we found a linear energy dependency of the scattering rate on binding energy, resembling the behaviour of a marginal-Fermi-liquid. Given the discussion above, It is relevant to note here that the top of the $\alpha_1$ band is very close to the Fermi level and will yield a van-Hove-singularity-like peak in the density of states. in close proximity to a van Hove singularity near the Fermi level. Following the argumentation of Ref. \onlinecite{Fink2015}, the presence of a van Hove singularity would induce non-Fermi-liquid behaviour for the quasiparticles.
The data presented here, therefore, can be taken to provide evidence for the importance of such phenomena in high-$T_c$ superconductors of iron parentage, besides the well-known case of the copper-oxides.~\cite{Chen2011, Varma1989}

\section{Conclusions}
In conclusion, using angle-resolved photoelectron spectroscopy (ARPES),  we have studied the electronic structure of Fe$_{1+y}$Te$_{1-x}$Se$_x$ superconductors.
From polarization-dependent measurements we disentangled the orbital character of the detected bands that are formed mainly by the combination of ${xz}$,  ${yz}$ and $z^2$ states in the vicinity of the Fermi level. 
We observed that the presence of excess Fe does not shift the bands in a rigid-band manner in these compounds.
The $k_z$ dependent band structure suggests weak Fermi surface warping along the $\Gamma - Z$ direction for the $\alpha_2$ band,  while  the $\alpha_1$ hole-like band that does not cross the Fermi level shows a moderate $k_z$ dispersion.
The mass enhancement factor ($m^*/m_b$) was not observed to change significantly from $\Gamma$ to $Z$ for the $\alpha_2$ band, but a dramatic change in $m^*/m_b$ was seen for the $\alpha_1$ band close to $k_z$=0.5 in units of $\pi/c$.
Despite this, near the $\Gamma$- and $Z$-points, both the $\alpha_1$ and $\alpha_2$ bands show the same mass enhancement factor within the range $m^*/m_b$=1.8$\pm$0.2. The observation of a $k_z$-independent Fermi velocity ($v_F$) for the $\alpha_2$ hole pocket is consistent with our DFT calculations.

We go on to show that the experimentally obtained imaginary part of the self-energy can be compared with both the marginal-Fermi-liquid and spin-fluctuation theoretical scenarios. 
The spin-fluctuation theory give a reasonable agreement to the data but with unrealistic parameters ($\lambda_{sf}$=7 and $\omega$=30 meV). 
The marginal-Fermi-liquid approach fitted the self-energy data for the $\alpha_1$ band well, yielding a coupling constant, $\lambda_{MFL}$=1.5, which is in close agreement with analogous constants derived for doped BaFe$_2$As$_2$ and NaFeAs iron pnictides.~\cite{Fink2015}
We discuss that the observed non-Fermi-liquid behaviour for the quasiparticles near the zone center in the 11 compounds could follow from the proximity of a van Hove singularity due to the $\alpha_1$ band to the Fermi level, thus making a direct link between the existence of a near $E_F$ van Hove singularity, non-Fermi-liquid behavior and high-$T_c$ superconductivity in iron-based compounds.

\section{Acknowledgements}
T.S. acknowledges support by the Department of Science and Technology (DST) through INSPIRE-Faculty program (Grant number: IFA-14 PH-86). T.S. thanks D. D. Sarma for his enormous support in I.I.Sc.  J.F. and I.E. acknowledge support by the German Research Foundation (DFG) through the priority program SPP1458. This work is a part of the research program of the Stichting voor Fundamenteel Onderzoek der Materie (FOM), which is financially supported by the  Nederlandse Organisatie voor Wetenschappelijk Onderzoek (NWO). I.E. acknowledges the support by a Kazan (Volga Region) Federal University grant targeted at strengthening the university's competitiveness in the global research and educational environment. The authors from CSIR-NPL would like to acknowledge financial support from the Govt. of India through the DAE-SRC outstanding researcher award scheme.

\bibliography{FeTe}

\end{document}